\documentclass[journal,12pt,onecolumn,draftclsnofoot]{IEEEtran}
\usepackage{amsmath,amsfonts}
\usepackage{amssymb}
\usepackage{algorithmic}
\usepackage{array}
\usepackage[caption=false,font=normalsize,labelfont=sf,textfont=sf]{subfig}
\usepackage{textcomp}
\usepackage{stfloats}
\usepackage{url}
\usepackage{verbatim}
\usepackage{graphicx}
\usepackage{comment}
\usepackage{cite}
\usepackage{xcolor}
\usepackage{algorithm}
\usepackage{subcaption}
\usepackage{hyperref}
\usepackage{caption}
\def\BibTeX{{\rm B\kern-.05em{\sc i\kern-.025em b}\kern-.08em
    T\kern-.1667em\lower.7ex\hbox{E}\kern-.125emX}}
\usepackage{balance}
\begin{document}
\title{Properties of Array Factor of Reconfigurable Intelligent Surfaces for Radio Communications}

\author{Symeon~Vaidanis,~\IEEEmembership{Graduate Student Member,~IEEE,}
        and~Athanasios~D.~Panagopoulos,~\IEEEmembership{Senior Member,~IEEE}
\thanks{S. Vaidanis is with the Communication Systems Department of EURECOM, Sophia Antipolis, 06410 France, and A. D. Panagopoulos is with the School of Electrical and Computer Engineering of National Technical University of Athens, 15780 Greece. E-mail: (see symeon.vaidanis@eurecom.fr, thpanag@ece.ntua.gr).}}

\maketitle

\begin{abstract}
This paper delves into the intricate aspects of Reconfigurable Intelligent Surfaces (RIS) in the antenna design. It investigates the properties of the array factor of RIS, highlighting the significance of phase shifts in establishing of the reflect path. Additionally, the paper investigates spatial positions requiring identical phase shifts, nullification of the array factor, the count of total maxima, second maximum determination (amplitude of the second lobe), and the calculation of the 3dB bandwidth. They are very useful for RIS-assisted radio communication systems. 
\end{abstract}

\section{Introduction}
Reconfigurable Intelligent Surfaces are thought as one of the most prominent technologies of the new 6G era and many applications of them have been proposed until now. The motivation of this paper is based on the issues of RIS as multiple access radio device \cite{VTtransaction:RIS_Multi_User} - \cite{IFStransaction:RIS_Multi_User} and the usage of RIS as a single radio device enhancing physical layer security \cite{IFStransaction:RIS_Physical_Layer_Security} - \cite{VTtransaction:RIS_Physical_Layer_Security}. Firstly, the array factor of RIS defines the coverage array of the RIS under some specific standards about the improvement provided by using RIS and about the interference levels from the side lobes of the RIS array factor. Secondly, the array factor plays a crucial role for physical layer security applications of the RIS because determines the indirectly maxima of the RIS array factor for a given configuration. These are possible  positions of the eavesdroppers (vulnerable points). 

\section{Path Loss Modeling and Array Factor}
We assume a passive RIS without inter coupling between its elements, which is located to the center of the coordination system, with M elements per row, which has the orientation of x axis, and N per column, which has the orientation of y axis. The spacing between the elements is $d_x$ and $d_y$ at the direction of rows and columns respectively. This spacing is an integer fraction of the wavelength $\lambda$ and it is expressed mathematically as $d_x = \frac{\lambda}{w_x} \:,\: d_y = \frac{\lambda}{w_y} \:,\: w_x,w_y \in \mathbb{N}$. We investigate a beamforming usage of RIS under far field conditions in which RIS is dedicated to a pair of transmitter (Tx) and receiver (Rx) and creates a reflect path, when the direct path is blocked. The channel model for this case has been proposed in paper \cite{WCtransaction:RIS_modeling} and the ration between the received power and the transmitted power can be described as following:
\begin{subequations}
    \begin{equation}
        \frac{P_r}{P_t} = \left | \sqrt{\frac{G^{t \rightarrow RIS \rightarrow r}}{LF^{t \rightarrow RIS \rightarrow r}}} \cdot M \cdot N \cdot  \frac{sinc\left(\frac{M \cdot u}{2}\right) \cdot sinc\left(\frac{N \cdot v}{2}\right)}{sinc\left(\frac{u}{2}\right) \cdot sinc\left(\frac{v}{2}\right)}  \right |^2
    \end{equation}
    \begin{equation}
        k \cdot \delta _1 \cdot \left( m - \frac{1}{2} \right) d_x + k \cdot \delta _2 \cdot  \left( n - \frac{1}{2} \right) d_y = \phi _{(n,m)}
    \end{equation}
    \begin{equation}
        u = k \cdot \left( \cos{\left(\gamma ^{RIS,t}_x\right)} +  \cos{\left(\gamma ^{RIS,r}_x\right)} + \delta _1 \right) \cdot d_x
    \end{equation}
    \begin{equation}
        v = k \cdot \left( \cos{\left(\gamma ^{RIS,t}_y\right)} + \cos{\left(\gamma ^{RIS,r}_y\right)} + \delta _2 \right) \cdot d_y
    \end{equation}
\end{subequations}
where $t$ denotes the transmitter, $r$ denotes the receiver, $G^{t \rightarrow RIS \rightarrow r}$ is the antenna gains of the transmitter, the incoming gain of RIS element, the outcoming gain of RIS element and the receiver, $LF^{t \rightarrow RIS \rightarrow r}$ is the double large scale fading losses including path loss and shadowing, $\phi _{(n,m)}$ is the phase shift of the $(m,n)$ element and $k$ is the propagation coefficient. Moreover, the $\cos{\left(\gamma _x\right)}$ and $\cos{\left(\gamma _y\right)}$ are equals to $\sin{\left(\theta \right)}\cos{\left(\phi \right)}$ and $\sin{\left(\theta \right)}\sin{\left(\phi \right)}$ respectively, where $\phi$ and $\theta$ are the azimuth and vertical angle respectively. Take into consideration the transformation between the Cartesian and spherical coordination systems, $\cos{\left(\gamma _x\right)}$ and $\cos{\left(\gamma _y\right)}$ are the projections to the x and y axis respectively of a radial scaling of the position vector of a point towards the unity sphere. Hence, $\cos{\left(\gamma ^{RIS,i}_x\right)} \equiv x_i \:,\: i\in \{t,r\} \:,\: x_i \in [-1,+1]$ and $\cos{\left(\gamma ^{RIS,i}_y\right)} \equiv y_i \:,\: i\in \{t,r\} \:,\: y_i \in [-1,+1]$. Furthermore, the array factor of the RIS can be defined as following :
\begin{equation}
    S(u,v) = M \cdot N \cdot \frac{sinc\left(\frac{M \cdot u}{2}\right) \cdot sinc\left(\frac{N \cdot v}{2}\right)}{sinc\left(\frac{u}{2}\right) \cdot sinc\left(\frac{v}{2}\right)}
\end{equation}
As we can see, the array factor has two independent components, which can be described through the same function $g(x,Y) = Y \cdot \frac{sinc\left(\frac{Y \cdot x}{2}\right)}{sinc\left(\frac{x}{2}\right)} = \frac{sin\left(\frac{Y \cdot x}{2}\right)}{sin\left(\frac{x}{2}\right)} $ with different inputs.

\section{Total Maxima and Nullification of Array Factor}
In the beamforming case, the optimal value of the $u,v$ are zero in order to achieve the maximum possible received power through constructive contribution to the receiver of the electromagnetic waves from each RIS's element, and so $\delta_1 = -x_t -x_r \:,\: \delta_2 = -y_t -y_r \:,\: \delta_1,\delta_2 \in [-2,+2]$. For every acceptable values of $x_t,y_t$, the set of the possible position of receiver $x_r,y_r$, which can be served from a configuration of RIS $\delta_1,\delta_2$ and the array factor achieves the maximum value, can be described as: $\max\{-1,-1+\delta_1\} \leq x_r \leq \min\{1,1+\delta_1\}$ and $\max\{-1,-1+\delta_2\} \leq y_r \leq \min\{1,1+\delta_2\}$.

The nullification of the nominator occurs when $\frac{u}{k \cdot d_x} = \frac{\mu_1 \cdot w_x}{M} \:,\: \frac{v}{k \cdot d_y} = \frac{\mu_2 \cdot w_y}{N}$ with $\mu_1 , \mu_2 \in \mathbb{Z}$ and the nullification of the denominator $\frac{u}{k \cdot d_x} = \eta_1 \cdot w_x \:,\: \frac{v}{k \cdot d_y} = \eta_2 \cdot w_y$ with $\eta_1 . \eta_2 \in \mathbb{Z}$. When the nominator and denominator nullify at the same time, in other words $\mu_1 = M \cdot \eta_1 \:,\: \mu_2 = N \cdot \eta_2$, and by applying l' Hopital theorem, it turns out that the array factor becomes maximum. 
In order to investigate the additional maxima of array factor, we have to define $u = - \hat{\delta_1} + \delta_1 \:,\: x_t + x_r = - \hat{\delta_1} \:,\: \delta_1 \neq \hat{\delta_1}$ where $\delta_1$ is the existed configuration of RIS and $\hat{\delta_1}$ is the desired configuration for the pair transmitter and receiver. Similar formulation applies for $v$. In continue, as $\delta_1,\delta_2 \in [-2,+2]$, for every acceptable values of $\delta_1,\delta_2$ we have that the set of the configurations of the RIS, which are indirect implemented, can be described as $\max\{-2,-2-\eta_1 \cdot w_x\} \leq \hat{\delta_1} \leq \min\{2,2-\eta_1 \cdot w_x\}$ and $\max\{-2,-2-\eta_2 \cdot w_y\} \leq \hat{\delta_2} \leq \min\{2,2-\eta_2 \cdot w_y\}$. Hence, as the parameter $w$ increases, which means that the spacing between the elements increases, the number of RIS configuration, which achieves maxima indirectly, decreases.

\section{Side Lobes and Full Width At Half Maximum}
As we can observe from the expression of RIS's array factor, the component of $u,v$ describes separately and independently the x and y axis respectively. Hence, we can investigate the amplitude of the highest second lobe and the full width at half maximum separately for each axis and then compare them. Moreover, taking into consideration the antenna array theory and the channel modeling expression of RIS, we are going to use the absolute value of the array factor risen to the power of 2 as studying function. 

First of all, in order to calculate the position of the highest second lobe for the one axis, we should maximize the contribution of the other and then find the nearest nullification to zero of the first partial derivative of the studying function. In other words, for axis x the $\text{second lobe} = \left| N \cdot g(u_2,M) \right|^2$ where $\frac{\partial g(u,M)}{\partial u}\left|_{u=u_2}\right. = 0 \rightarrow M \cdot \tan \left(\frac{u_2}{2}\right) = \tan \left(\frac{M \cdot u_2}{2}\right)$ and $u_2$ is the closest nullification to zero. The same process with the profound changes can be followed in order to calculate the highest second lobe for the y axis. As we can see the ratio between the main and the second lobe is independent and constant for large number of elements, is not affected by the spacing and is equal to $13.25dB$. 
\begin{figure}[h]
\begin{center}
\noindent
  \includegraphics[width= 5in]{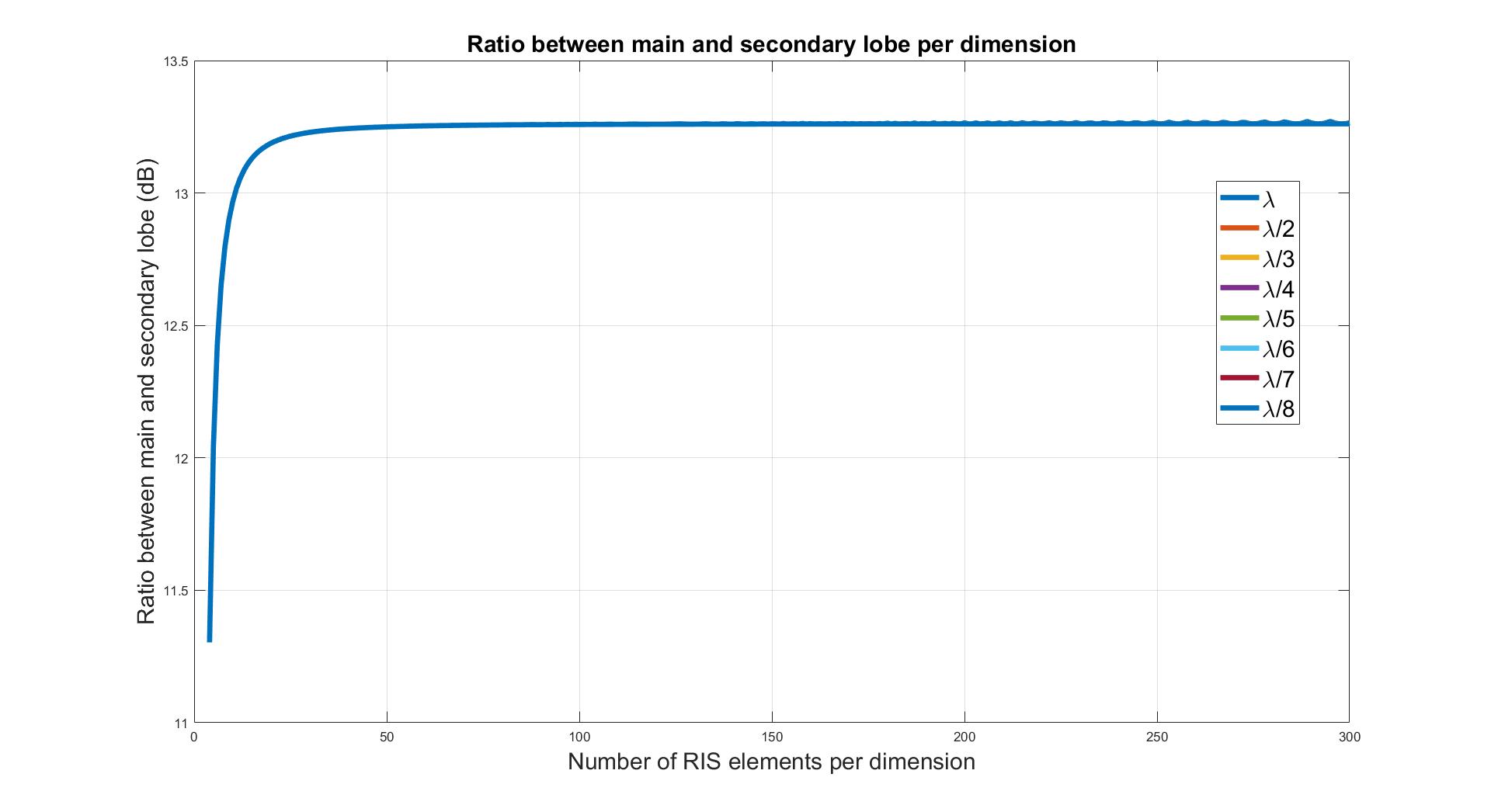}
  \caption{Side Lobe Amplitude}\label{Fig1Label}
\end{center}
\end{figure}
In continue, the Full Width at Half Maximum (FWHM) of the RIS's antenna factor can be calculated for each axis separately, while the contribution of the other axis is maximized. In particular, for axis x the $\Delta_{3dB} = \left|\frac{w_{x}\cdot u_{3dB}}{\pi}\right| \:,\: \left| g(u_{3dB},M) \right|^2 = \frac{M^2}{2}$. The same process with the profound changes can be followed in order to calculate the highest second lobe for the y axis. From the graph, we can extract the conclusions that as the number of RIS elements increases, the FWHM decreases as expected from antenna array theory and as the spacing between the element decreases, the FWHM increases. Moreover, we can observe that FWHM is inversely proportional to the number of elements.
\begin{figure}[h]
\begin{center}
\noindent
  \includegraphics[width=5in]{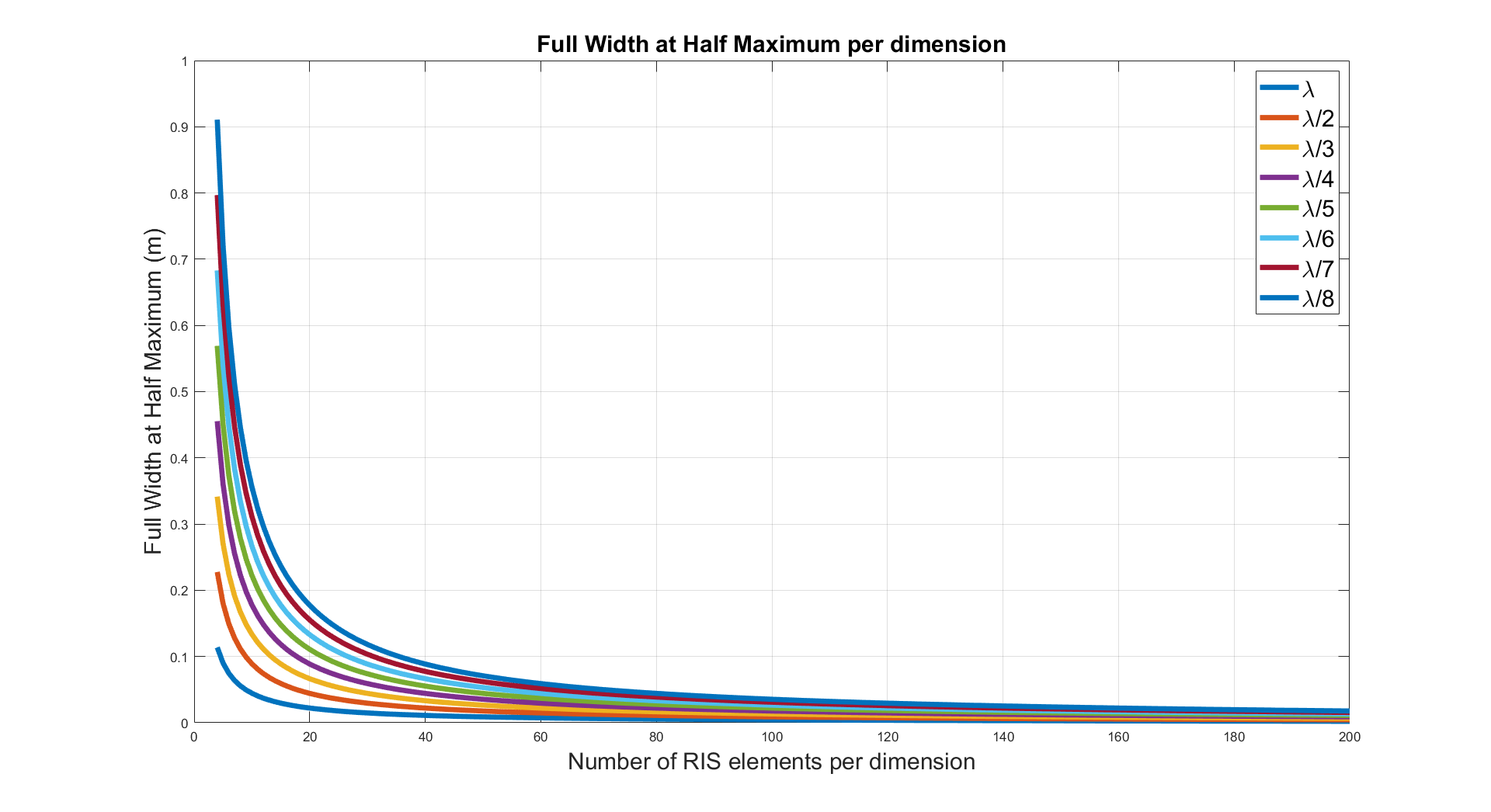}
  \caption{FWHM}\label{Fig2Label}
\end{center}
\end{figure}
\section*{Conclusions and Future Work}
In conclusion, the study of array factor of RIS reveals some important features of this technology, namely the indirect maxima for a certain configuration, the constant first to second lobe ratio for large number of elements, the significant decrease of FWHM by the increase of elements' number and the increase of FWHM by the decrease of the spacing between the elements. The study of the impact of small scale fading and the uncertainty to the position of transmitter and receiver to the RIS array factor are to prominent directions for future work.


\end{document}